To cite this paper, please use the following reference:

Tavakkol, S., & Zarrati, A. R. (2018). Wall and Bed Shear Force in Rectangular Open Channels. Scientia Iranica

# Wall and Bed Shear Force in Rectangular Open Channels

Sasan Tavakkol

PhD Candidate, Department of Civil and Environmental Engineering, University of Southern California, 920 Downey Way., Los Angeles, CA 90089, USA. Email:tavakkol@usc.edu (Corresponding Author)

Amir Reza Zarrati

Professor, Department of Civil and Environmental Engineering, Amirkabir University of Technology, 424 Hafez St., Tehran, Iran. Email: zarrati@aut.ac.ir

**Abstract**: A method is introduced to determine the percentage of the total shear force acting on the walls and bed in rectangular open channels. The proposed method takes both the velocity gradients and secondary currents into account. In the current method the channel cross section is divided into subsections using the bisectors along which there are no secondary flow effects, and isovels orthogonal trajectories along which there is no shear stress. Based on these subsections and assuming the equilibrium between the shear force and gravitational force, the share of the bed and wall from the mean shear force of the flow are calculated. Calculated wall and bed shear forces are in very good agreement with experimental data with an average relative error less than 5%. It is also shown that neglecting the effect of secondary currents and only assuming zero-shear division lines does not yield to acceptable results. The method also provides a possible range for wall and bed shear forces which nicely covers the experimental data. In addition, given the location of the maximum velocity, the method is used for steady uniform flows, carrying suspended sediment with good agreement.

1. Introduction

In a uniform flow, the total shear force acting on the wetted perimeter can be calculated in terms of the bed slope, the hydraulic radius, and the fluid density. However, calculating the portion of the shear force acting on side walls and the channel bed is an object of research even after decades [1-7]. Calculation of the shear force acting on bed and channel walls is important in hydraulic engineering studies. For example the bed mean shear stress is needed to calculate the bed load, and the wall mean shear stress is needed in studying bank erosion, channel migration and river morphology. Moreover, the knowledge of shear force ratio is needed in laboratory flume studies. Empirical results from flume studies are often subject to sidewall friction effects, a procedure based on shear stress ratio can be used to remove these effects, which is referred to as sidewall correction procedure [8-9]. The present paper introduces a method for determining the percentage of the total shear force acting on the walls *(%SF$_w$)* or bed *(%SF$_b$)* based on the location of the maximum velocity on the centerline of a rectangular channel cross section. Having the *%SF$_w$* and *%SF$_b$*, one can calculate the shear distribution on channel boundary [10].

Among different methods for local shear stress calculation, there are a series of simple and popular methods that rely on the splitting of the channel cross-section in different subsections. In these methods it is assumed that the weight component of the fluid inside each subsection is balanced by the shear force along the corresponding wetted perimeter. Leighly (1932) [11] employed this idea for the first time to estimate the shear stress distribution. His work is recently highlighted by Buffin-Bélanger (2010) [12]. Leighly mapped the channel cross section by lines of constant velocity (isovels) and their orthogonal curves. He then defined the subsections as the area between two consecutive orthogonal curves. Since the velocity gradient across orthogonal curves is zero, these curves are assumed to be "surfaces of zero shear"; however, presence of secondary flows cause momentum exchange and consequently resistance force along these surfaces. The interaction among main flow, secondary flows and shear stress distribution is well investigated by Chiu and Chiou (1986) [13] and Yang (2005) [14].

After Leighly (1932), several researchers including Schlichting (1936), Keulegan (1938), Haywood (1940), Johnson (1942), and Einstein (1942) [15-19] developed various methods for predicting shear stress distribution based on dividing the channel cross section assuming zero-shear division lines. According to this assumption the cross section of a channel can be separated by two zero-shear division lines into three parts (Fig. 1). In this assumption, the component of the fluid weight in each area is assumed to be balanced by the resistance of the channel boundaries in that area. Yang and Lim (1997) [1] proposed that the surplus energy of a unit volume of fluid in a three-dimensional channel is dissipated at the "closest" wall boundary. Based on this assumption they developed a linear implicit equation for the zero shear division lines.

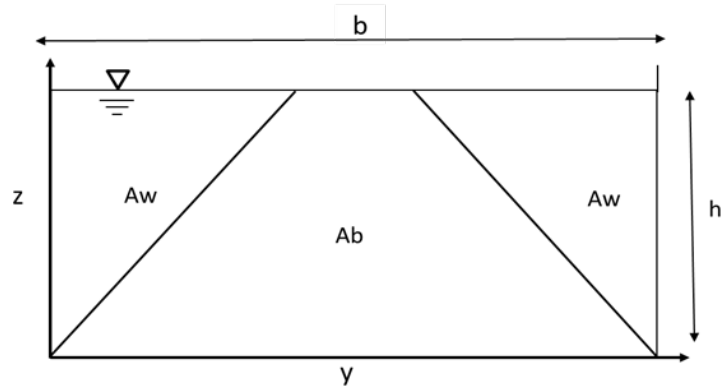

Fig. 1. Cross-section of a channel divided into bed ($A_b$) and wall areas ($A_w$).

Later, Gou and Julien (2002, 2005) [7, 2] employed the differential form of momentum and mass conservation equations to formulate shear force distribution between walls and bed of a rectangular channel. Deriving this formulation is straightforward using the integral form of conservation equations. Employing the Schwarz–Christoffel transformation, they formulated the orthogonal curve to isovels drawn from the channel bottom corner and assumed it as a zero-shear division line. Their proposed equation, as stated by the authors, did capture the dip phenomenon and implied the maximum velocity to occur at the water surface. The assumed curve as an orthogonal trajectory did not conform with the experimental data of Nezu and Nakagawa (1993) [20]. Therefore, Gou and Julien (2005) [2] introduced two lumped empirical correction factors to correct the error caused by neglecting the secondary currents and deviation of assumed division line curve from the reality. The shape of the division lines (curved or linear), were questioned by Yang and Lim (2006).

In the present paper, first, the importance of different terms of momentum equation on a control volume surrounded by an isovel and a control volume between two curves perpendicular to the isovels are explained. The possibility of neglecting the secondary currents is examined next, and it is shown that assuming zero-shear division lines is not acceptable. Finally a new method is introduced for calculating the percentage of shear force acting on the channel bed and side walls. The proposed subdividing method is applied on uniform rectangular open channels with fully developed turbulent flows. However the general concept can be extended to other cross-sectional shapes.

## 2. Theoretical Consideration

For an arbitrary control volume on the cross section of a rectangular channel (Fig. 2) with assuming a steady uniform flow, the continuity and momentum equations can be written as:

$$\oint \rho(\mathbf{V}.\mathbf{n})dl = 0 \qquad (1)$$

$$\rho g A S + \oint \tau_{nx} dl - \oint \rho u(\mathbf{V}.\mathbf{n})dl = 0 \qquad (2)$$

where $\rho$=density; $A$= area of the control surface on y-z plane; $g$=gravitational acceleration; $S$=channel slope; $\mathbf{V}$=velocity component of flow on y-z plane; $\mathbf{n}$=unit normal vector pointing outside of the control volume; $l$=length along the boundary of surface; $\tau_{nx}$=shear stress in the flow direction x applying on the plane perpendicular to $\mathbf{n}$. The flow velocity components, $u$, $v$, and $w$ are in longitudinal, lateral, and vertical or in x, y, and z directions, respectively.

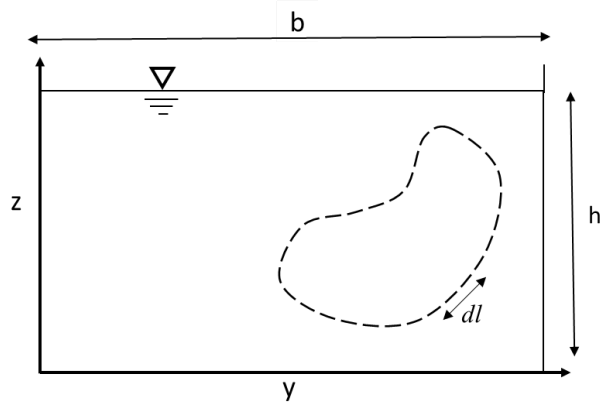

**Fig. 2. An arbitrary control volume in a channel cross section.**

For a control volume surrounded by an isovel, the momentum equation can be written as

$$\rho g A S + \oint \tau_{nx} dl - u \oint \rho(\mathbf{V}.\mathbf{n})dl = 0 \qquad (3)$$

Considering Eq. (1), the third term of Eq. (3), which represents the effect of secondary flows, vanishes and Eq. (3) reduces to:

$$\rho g A S + \oint \tau_{nx} dl = 0 \qquad (4)$$

Eq. (4) shows that the gravity component of a control volume limited to an isovel is balanced only by the shear stress acting on its surface due to velocity gradients. On the other hand, if the momentum equation is applied to a control volume surrounded by two curves (polygon MNOM in Fig. 3) which are orthogonal to isovels (Fig. 3) one can write:

$$\rho g A_{MNOM} S + \oint_{MNOM} \tau_{nx} dl - \oint_{MNOM} \rho u (V.n) dl = 0 \qquad (5)$$

Where $A_{MNOM}$ is the area of the control volume surrounded by the polygon *MNOM* and both integrals are calculated on the same polygon. The first integral on the left-hand side of Eq. (5) can be calculated on subsections of the polygon *MNOM* as

$$\int_{MNOM} \tau_{nx} dl = \int_{MN} \tau_{nx} dl + \int_{NO} \tau_{nx} dl + \int_{OM} \tau_{nx} dl \qquad (6)$$

$\tau_{nx}$ can be assumed as

$$\tau_{nx} = \rho (v_t + v) \frac{\partial u}{\partial n} \qquad (7)$$

where $v$ and $v_t$ are kinematic and turbulence viscosity of water, respectively. Since NO and OM are orthogonal to isovels, $\partial u/\partial n$ is zero along them; therefore the integrals $\int_{NO}$ and $\int_{OM}$ are zero. Considering the average shear stress on MN as $\tau_{MN}$, and its length as $\delta P_{MN}$, and Eq. (5) we can derive:

$$|\tau_{MN}| = \frac{1}{\delta P_{MN}} (\rho g A_{MNOM} S - \oint_{MNOM} \rho u (V.n)) dl \qquad (8)$$

For a known 3D velocity field one can calculate the shear stress distribution on channel boundary employing Eq. (8); but usually all 3 velocity components are not known. Note that viscosity does not show up in Eq. (8), but its effects is considered through the velocity field.

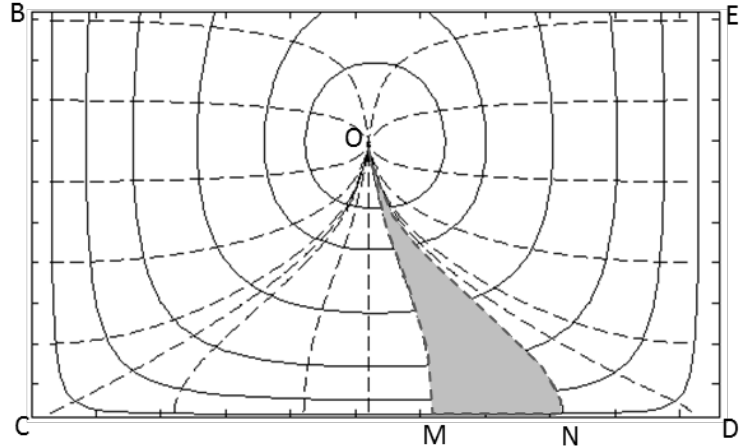

Fig. 3. Isovels and orthogonal curves for *b/h*=2 (Isovels are calculated following Chiu and Hsu, 2006).

Based on the above discussion, shear resistance cannot be neglected neither on a control volume surrounded by isovels nor surrounded by orthogonal curves to isovels. Thus it seems unlikely to find a series of zero shear curves in the cross section of an open channel. The channel centerline however is an exception on which both the shear resistance due to secondary currents and velocity gradient are zero. If secondary currents are neglected, despite their indisputable effect, Eq. (8) becomes:

$$|\tau_{MN}| = \frac{\rho g A_{MNOM} S}{\delta P_{MN}} \tag{9}$$

Eq. (9) describes Leighly's assumption. Guo and Julien (2005)[2] employed the same assumption to propose a first approximation for mean shear stress on channel bed and walls.

Even with neglecting the secondary currents, the distribution of the main velocity, *u*, should be known to calculate the curves orthogonal to isovels and from that the distribution of shear stress on the channel boundary. It should also be mentioned that in an open channel the maximum velocity often occurs at a point below the water surface (Henderson, 1966)[22]. Knowing the location of the maximum velocity is important because it is the point of concurrency of the isovel orthogonal trajectories.

Several researchers proposed methods to calculate the velocity distribution in an open channel [22-26]. Maghrebi and Rahimpour's [25] method similar to Houjou et al. (1990) [23] predicts the location of the maximum velocity below water surface for only *b/h*<2, where b is the channel width and h is the flow depth. It seems that the results of these methods are due to the fact that an open channel with *b/h*=2 is half of a square conduit in which the velocity profile is symmetrical in the y and z directions [23]. Cacqueray et al. (2009) [3] who examined the formulation of Guo and Julien (2005) [2] using detailed computational fluid dynamic simulation, reported that the maximum velocity occurs on the flow surface for a channel with *b/h*=2 too. However, it is widely reported that the dip phenomenon is observed for channels with *b/h*<5 [27-31].

The velocity distribution proposed by Chiu and Hsu (2006) [26] is employed in the present paper to calculate shear force distribution. Chiu and Hsu (2006) presented a probabilistic approach to calculate the velocity distribution following Chiu (1987), Chiu and Murray (1992), Chiu and Said (1995), and Chiu and Tung (2002) [32-35]. This method relies on a parameter which could be found according to various hydraulic parameters of channel such as the ratio of average velocity to the maximum velocity and the location of the maximum velocity. The location of maximum velocity can be calculated from Yang et al. (2004) [24] as:

$$\frac{\varepsilon}{h} = 1 - \frac{1}{1 + 1.3e^{-b/2h}} \tag{10}$$

in which $\varepsilon$ is the depth of the maximum velocity.

The isovels and their orthogonal curves for a channel with b/h=2 calculated by Chiu and Hsu (2006)'s [26] method based on Eq. (10), is shown in Fig. 3. The slope of isovel orthogonal trajectories at any point can be calculated from $\tan(\theta)=(\partial u/\partial z)/(\partial u/\partial y)$. Starting from a certain point on the boundary and following the slope, the orthogonal trajectories can be drawn. If the momentum exchange generated by the secondary currents on isovel orthogonal curves is neglected, the gravity component of the mass in surrounded by the closed polygon BOEB (Fig. 3), remains unbalanced which leads to underestimation of the total shear force on the channel boundary.

Fig. 4 compares the estimation of Eq. (9) for $\%SF_b$ with the empirical equation presented by Knight et al. (1984) [36]. As it is illustrated in Fig. 4, neglecting the secondary currents leads to underestimating the bed shear force and is not acceptable. The location of maximum velocity moves toward the flow surface in wider channels and therefore predictions of Eq. (9) becomes more accurate for larger *b/h*.

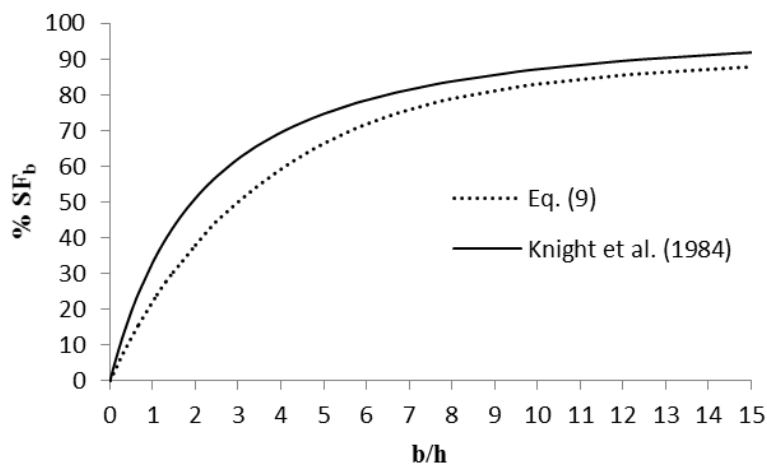

**Fig. 4.** Comparison of *%SF$_b$* predicted by neglecting secondary currents with the empirical equation presented by Knight et al. (1984).

### 3. The Present Methodology

Eq. (8) can be employed to calculate the mean shear on the bed and side walls of an open channel if the effect of secondary currents can be calculated. If the orthogonal trajectories to the isovels are drawn from the channel corners, they will intersect on the maximum velocity location (P$_2$ curves in Fig. 5). The weight component of water in the area limited to P$_2$ curves and channel bed, *A$_{bmin}$*, must be less than the shear force acting on bed. This is since P$_2$ curves are orthogonal to isovels, with no velocity gradients and consequently no shear stress acting on them. But there are secondary flows passing across these curves which cause momentum exchange. Secondary currents enter this area near the channel centerline (Henderson 1966) [22], where the main flow velocity is maximum. Due to continuity, the flow which enters this area near the channel centerline leaves this area at locations where flow velocity is smaller. It can therefore be said that the acting force from momentum transfer into the mass of this area is in the flow direction. It will later be shown that the measured bed shear force is in excess of the weight component of water in this area (*A$_{bmin}$*).

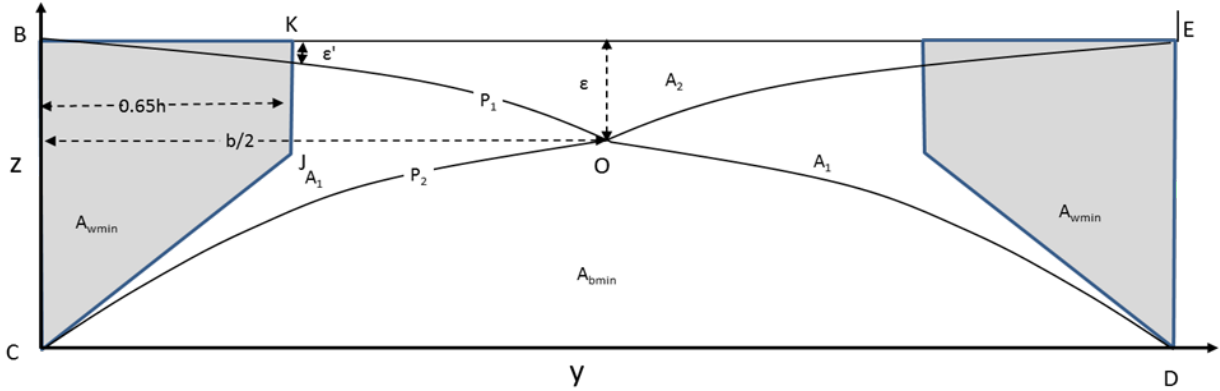

Fig. 5. Dividing the channel cross section by bisectors (CJ) and isovels orthogonal curves ($P_1$ and $P_2$).

Gessner (1964) [37] suggested that the secondary flow cells in the corners of rectangular ducts are divided by the corner bisector. Experimental data presented by Tominaga et al. (1989) [38] show that secondary flows are strong in a region within a distance equal to $0.65h$ from channel bottom corners. Zheng and Jin (1998) [39] and Jin et al.(2004) [40] assumed the same structure for corner secondary currents as Gessner (1964) [37] in rectangular open channels, with the boundary of $y/h =0.65$ and $z/h =0.65$. It is assumed that the secondary flow cells have negligible effect on the flow mechanism out of this boundary [40]. The effective area of secondary currents on the walls side can be therefore limited to the sidewall, bisector and a vertical line at $y/h =0.65$ (region CJKBC, Fig. 5). The weight component of water in this region must be less than the wall shear force. This is since it is assumed that secondary flows do not pass through the boundaries of this area, and therefore there is no momentum exchange due to secondary currents across these boundaries. However there could be momentum exchange due to shear force in the flow. Since in general, flow velocity decreases toward the sidewall the shear force acting on the CJK is in the flow direction and therefore, its value is added to the weight component of CJKBC. This area, which is called $A_{wmin}$, can be easily calculated as:

$$A_{w\min} = 2\times(\frac{h+0.35h}{2}\times 0.65h) = 0.65(1.35h^2) \approx 0.8h^2 \quad \text{for } b/h>1.3 \quad (11)$$

In channels with $b/h< 1.3$ the corner bisectors intersect below $z=0.65h$. This case will be discussed later.

According to these assumptions, CJK and $P_2$ in Fig. 5, determine the minimum share of sidewalls and bed from total shear force, respectively. The weight component of the area between these curves is assumed to be balanced partially by shear force acting on the sidewalls and partially with the shear force acting on the bed. The shared areas are divided into $A_1$ and $A_2$ by $P_1$ curve (Fig. 5). $P_1$ is the orthogonal curve to isovels drawn from upper corners of the channel cross section. Consider the share of side walls from weight component of $A_1$ and $A_2$, to be $\eta_1$ and $\eta_2$, respectively. One can therefore calculate $\%SF_w$ and $\%SF_b$ as below

$$\%SF_W = \frac{2\tau_w h}{\rho g A S}100\% = \frac{100\%}{A_t}(A_{w\min} + \eta_1 A_1 + \eta_2 A_2) \quad (12)$$

$$\%SF_b = \frac{\tau_b b}{\rho g A S}100\% = \frac{100\%}{A_t}(A_{b\min} + (1-\eta_1)A_1 + (1-\eta_2)A_2) \quad (13)$$

where $\tau_b$ and $\tau_w$ are mean bed and wall shear stresses, respectively, and $A_t$ is the channel cross sectional area. To determine $\eta_1$ and $\eta_2$, $A_1$ and $A_2$ must be calculated first. To calculate these areas, equations for $P_1$ and $P_2$ must be known. $P_1$ can be approximated by a parabola. $P_1$ passes through $(0,h)$ and $(b/2, h-\varepsilon)$. It is also perpendicular to the channel sidewall as an isovel; therefore its derivative is zero in $(0,h)$. The equation of $P_1$ can therefore be written as:

$$z = -\frac{\varepsilon}{(b/2)^2} y^2 + h \quad (14)$$

Similarly, $P_2$ passes through $(0,0)$ and $(b/2, h-\varepsilon)$. At the corner, $P_2$ must be perpendicular both to the channel bed and wall which is not possible. To remove this paradox, following Chiu and Chiou (1986) [13] the channel corner is assumed as a curve and therefore the slope of $P_2$ at the corner is assumed equal to 1.

Contrary to $P_1$, $P_2$ cannot be approximated by a parabola. A parabolic approximation significantly deviates from $P_2$ for larger $b/h$ and even may intersect water surface. To avoid this problem $P_2$ is approximated by a parabola starting on channel corner and a straight line following the parabola at a point, say $(y_i, z_i)$. Considering these boundary conditions $P_2$ can be expressed as:

$$z = a'y^2 + b'y + c' \qquad\qquad y \leq y_i \qquad\qquad (15.a)$$

$$z = a''y + b'' \qquad\qquad y > y_i \qquad\qquad (15.b)$$

where

$$a' = \frac{z_i - y_i}{y_i^2} \quad ; \quad b' = 1 \quad ; \quad c' = 0$$

$$a'' = \frac{h - \varepsilon - z_i}{b/2 - y_i} \quad ; \quad b'' = h - \varepsilon - a''\frac{b}{2}$$

The slope of the parabola and the line are equal on their intersection point. This condition results in the following equation between $y_i$ and $z_i$:

$$y_i^2 - \left(z_i + \frac{b}{2} + (h - \varepsilon)\right)y_i + z_i b = 0 \qquad\qquad (16)$$

For a given $z_i$, one can calculate $y_i$ from Eq. (16) and consequently the unknown coefficients $a'$, $a''$, and $b''$. $z_i$ is found by employing least squares regression between Eq. (15) and isovel orthogonal curves obtained based on Chiu and Hsu (2006) [26] and Eq.(10). Plotting $z_i/(h-\varepsilon)$ against $b/h$ indicates that $z_i/(h-\varepsilon)$ is almost constant and can be assumed equal to 0.93 for different values of $b/h$ with about 2% averaged relative error. Fig. 6 compares Eq. (15) with the mentioned numerical data assuming $z_i/(h-\varepsilon)=0.93$ with good agreement.

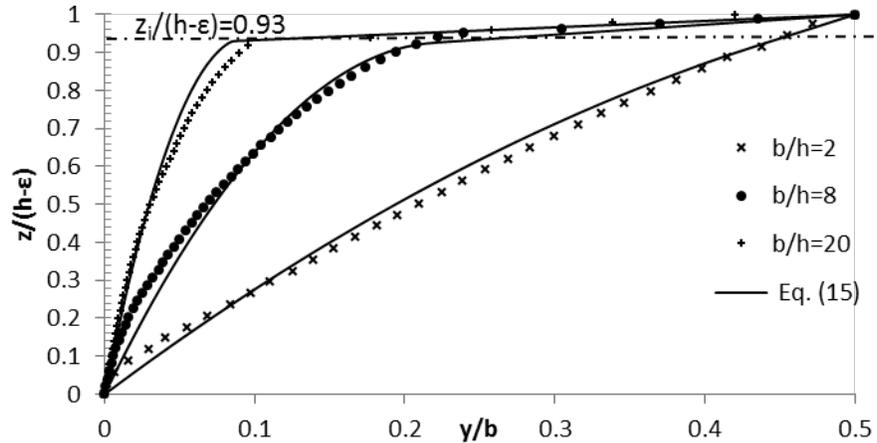

**Fig. 6. Comparison of Eq. (15) for $z_i/(h - \varepsilon) = 0.93$ with numerical data calculated from Chiu and Hsu (2006) method.**

According to the equations found for $P_1$ and $P_2$, $A_1$ and $A_2$ can be presented as:

$$A_1 = bh - A_{w\min} - A_{b\min} - A_2 \tag{17}$$

$$A_{b\min} = 2\left(\frac{a'y_i^3}{3} + \frac{y_i^2}{2} + \frac{z_i + h - \varepsilon}{2}(\frac{b}{2} - y_i)\right) \tag{18}$$

$$A_2 = 2\left(\frac{b}{6}(\varepsilon - \varepsilon')\right) = \frac{b}{3}(\varepsilon - \varepsilon') \tag{19}$$

where $\varepsilon'$ is the vertical distance between the intersection point of $P_1$ and CJK from water surface (Fig. 5). $\varepsilon'$, which is shown on Fig. 5, can be calculated from Eq. (16) as:

$$\varepsilon' = \frac{\varepsilon}{(b/2)^2}(0.65h)^2 \tag{20}$$

The theory can be extended to $b/h<1.3$ with considering necessary changes in the equations. For a channel with $b/h<1.3$, the bottom corner bisectors intersect the channel axis in a point like N and two different patterns may occur (Fig. 7). The maximum velocity may be below point N (Fig. 7 a), which means $h-\varepsilon<b/2<0.65h$. This pattern is similar to the pattern of channel cross section with $b/h>1.3$, but Eq. (11) and (19) must be substitute by $A_{w\min}=b(h-b/4)$ and $A_2 \approx 0$, respectively. The maximum velocity may occur above the point N, which means $b/2<h-\varepsilon<0.65h$ (Fig. 7 b). In this case, $A_{w\min}$ is limited to $P_2$ and side wall and $A_{b\min}$ is limited to the bisectors and channel bed and they can therefore be easily calculated.

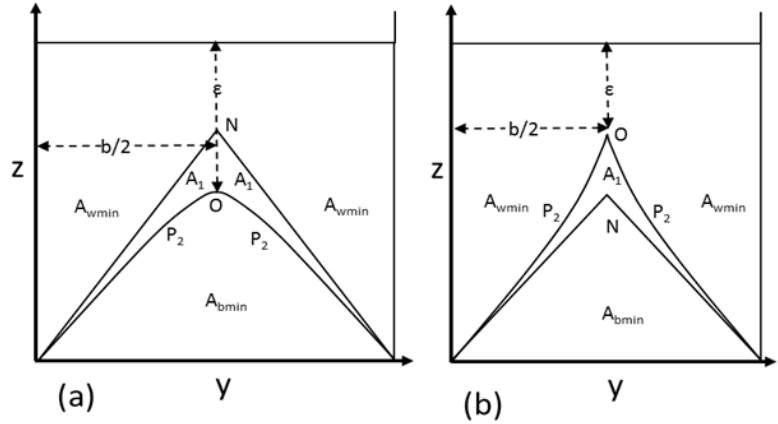

**Fig. 7. Possible cross-sectional division lines for *b/h*<1.3**

Employing Eq. (10) to calculate $\varepsilon$, shows that in *b*/*h*=1.16 point N conforms to point O. In this special and interesting case, the bottom corner bisectors and P$_2$ conforms and therefore can be assumed as actual division lines across which there is no shear stress due to velocity gradient or momentum exchange due to secondary currents. In this case $\eta_1$ and $\eta_2$ are eliminated from Eq. (12) and *%SF$_w$* is equal to %71. Fig. 8 compares this value with experimental results from different researches for *b*/*h* values close to 1.16. As can be seen from this figure the *%SF$_w$* calculated from the present theory is surrounded by the experimental results. A Gaussian weighted average from the shown experimental results on Fig. 8 gives *%SF$_w$*=69% for *b*/*h*=1.16 which is very close to predicted value of the present model. However, estimation of Knight et al. (1984)'s empirical equation and Guo and Julien (2005) [2] is 63.2% and 62.9%, respectively.

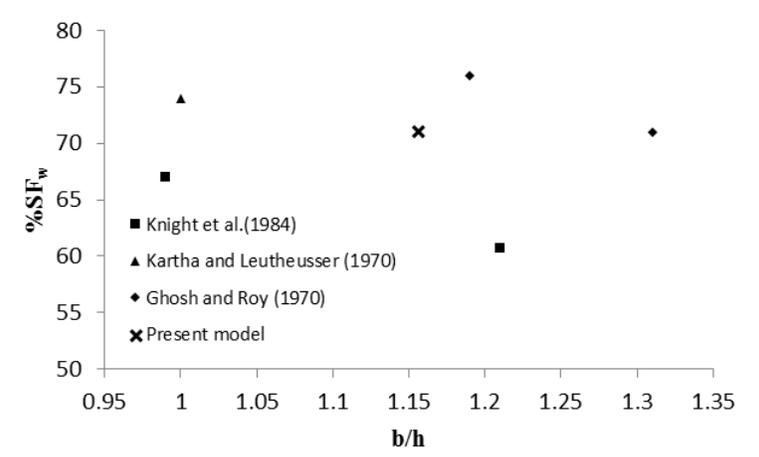

**Fig. 8. Comparison of *%SF$_w$* for *b/h*=1.16 with experimental results.**

## 4. Calibration

Fig. 9 compares the result of Eq. (12) with experimental data presented by Knight et al. (1984) [36] from various researchers [41-47]. For $\eta_1=\eta_2=0$, Eq. (12) gives the minimum possible value of $\%SF_w$ and for $\eta_1=\eta_2=1$, it gives the maximum possible value of $\%SF_w$. It can be seen in Fig. 9, that the experimental data lies very well between the predicted minimum and maximum values for $\%SF_w$ which demonstrates the validity of dividing the channel sections in to $A_{bmin}$ and $A_{wmin}$. Contrary to coefficients used in previous related works, the coefficients $\eta_1$ and $\eta_2$ are physically meaningful and limited to a range between 0 and 1. In addition, considering the maximum and minimum values of $\%SF_w$ illustrated in Fig. 9, it can be seen that the present model is not significantly sensitive to the values of $\eta_1$ and $\eta_2$.

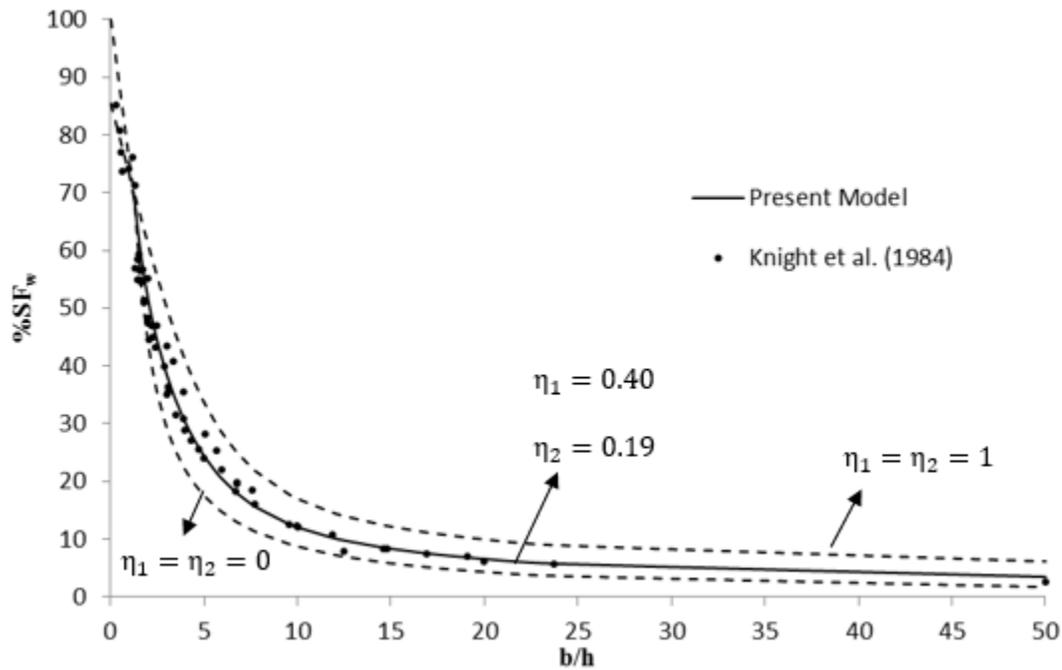

**Fig. 9. Comparison of the present model results for $\%SF_w$ with experimental data from various researchers (Knight el al, 1984).**

Employing the least squares method to estimate the share coefficients results in $\eta_1=0.40$ and $\eta_2=0.19$ with the coefficient of determination of $R^2=0.98$. According to these obtained values for $\eta_1$ and $\eta_2$, $\%SF_w$ can be calculated from:

$$\%SF_w = \frac{100\%}{bh}(A_{w\min} + 0.40A_1 + 0.19A_2) \tag{21}$$

where $A_{wmin}$, $A_1$, and $A_2$ are calculated from Eq. (11), Eq. (17), and Eq. (19), respectively. $\%SF_w$ values can be also read from Fig. 10 given the values of $b/h$ and $\varepsilon/h$.

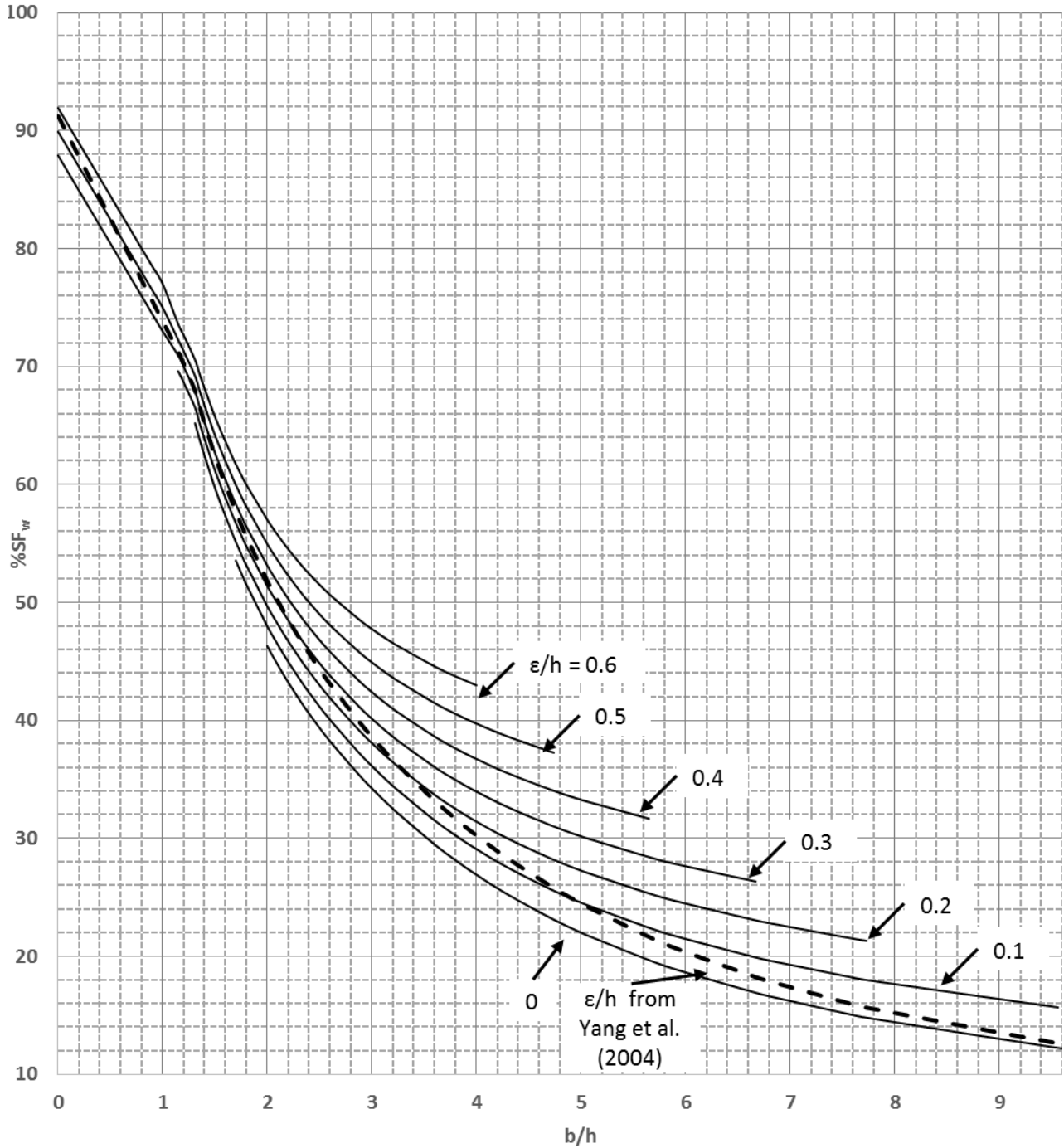

Fig. 10. Values of $\%SF_w$ for different values of b/h and ε/h.

## 5. Validation

Fig.11 compares the present model (Eq. 21) with experimental data of wall shear force from Seckin et al. (2006) [48] which are neither included in the calibration of the present model, nor in calibration of Gou and Julien (2005) [2] model. Considering the relative difference, RD, as:

$$RD = \left| \frac{measured - calculated}{measured} \right| \times 100$$

and excluding the experiment with $b/h$=13.18 which is out of range, the average relative difference between the experimental data and predicted values by present model is 4.7%, while it is 7.7% for Gou and Julien (2005) [2]. Eq. (21) is therefore, in very good agreement with the experimental data.

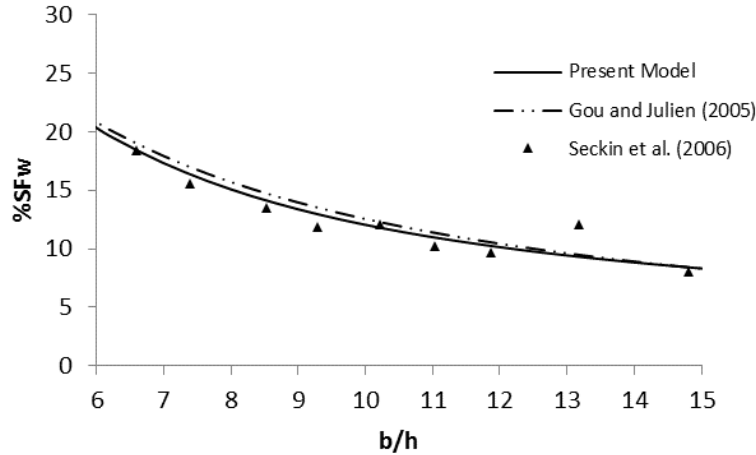

**Fig.11. Comparison of present model for $\%SF_w$ with unseen experimental data.**

Results of comparison of the present model (Eq. 21) and Gou and Julien (2005) with experimental measurements of Xie (1998) and Chanson (2000) for $\%SF_w$ are shown in Table 1. The relative error for the present model is 8.8% while it is 9.6% for Gou and Julien (2005).

**Table 1.** Comparison of present model and Gou and Julien (2005) (G&J) for $\%SF_w$ with experimental data of Xie (1998) and Chanson (2000)

|  | b/h | Measured | G&J | Eq. (21) |
|---|---|---|---|---|
| Xie (1998) | 3.2 | 36 | 36.40 | 36.62 |
| Chanson (2000) | 2.9 | 38 | 39.26 | 39.62 |
| | 3.24 | 37 | 36.05 | 36.25 |
| | 5.32 | 19 | 23.35 | 22.98 |
| | 5.95 | 18 | 20.99 | 20.50 |

Table 2 compares the result of Eq. (21) with experimental data of average bed shear velocity from experimental data of Coleman (1986), Lyn (1986), 2000, Muste and Patel (1997) [51-54], and predicted values by Gou and Julien (2005) [2]. The average bed shear velocity is calculated from $u_{*b} = \sqrt{\tau_b/\rho}$. In this table, $u_{*b,m}$ represents the measured values, $u_{*b,21}$ is the results of Eq. (21) and calculating $\varepsilon$ from Eq. (10) and $u_{*b,G\&J}$ is the prediction of Gou and Julien (2005) [2]. It can be seen that the predicted values by the analytical models are slightly smaller than the measured values. This is due to the fact that the measured values are on channel centerline, but the predicated values are mean bed shear velocities. The average relative difference between results of Eq. (21) and experimental data is 3.2% while it is 3.5% for Gou and Julien (2005).

**Table 2. Comparison of present model for bed shear velocities (cm/s) with experimental data**

|  |  | $S \times 10^3$ | b/h | $u_{*b,m}$ | $u_{*b,21}$ | $u_{*b,G\&J}$ |
|---|---|---|---|---|---|---|
| Coleman (1986) | Run 1 | 2 | 2.07 | 4.1 | 4.08 | 4.14 |
| Lyn (1986, 2000) | C1 | 2.06 | 4.08 | 3.1 | 3.05 | 3.05 |
|  | C2 | 2.7 | 4.09 | 3.7 | 3.49 | 3.49 |
|  | C3 | 2.96 | 4.64 | 3.6 | 3.51 | 3.5 |
|  | C4 | 4.01 | 4.69 | 4.3 | 4.07 | 4.06 |
| Muste and Patel (1997) | CW01 | 0.739 | 7 | 2.92 | 2.79 | 2.78 |
|  | CW02 | 0.768 | 7.1 | 2.92 | 2.83 | 2.82 |
|  | CW03 | 0.813 | 7.16 | 2.98 | 2.9 | 2.89 |

Coleman (1981, 1986) [51, 55] reported experiments conducted in a channel with $S=2\times10^3$, $b=35.6$ *mm*, and $Q=0.64$ *m³/s*, where $S$ was the bottom slope and $Q$ was the channel discharge. The experiments were carried out to investigate the effect of sediment suspension on velocity profile. Coleman reported the location of maximum velocity in all experiments which can therefore be used directly in Eq. (21).

Table 3 compares the results calculated from the present method and Guo and Julien (2005) with experimental data of Coleman (1986). In this table $u_{*b,21}$ is the result of Eq. (21) by directly substituting the $\varepsilon$ from experiment. The total amount of sand suspended in the flow is given in this table in terms of Kg as $M_s$. Moreover the average volumetric suspended sediment concentration ($\varphi$) of flow is calculated from available data in Coleman (1986) and provided in the table. It can be seen that the present model is predicting the average shear velocities equal or slightly smaller than $u_{*b,m}$; however, $u_{*b,G\&J}$ is always larger than $u_{*b,m}$. The predicted mean values, as mentioned before, should be slightly smaller than the measured local values since the measurements were made at the channel centreline. The measured shear velocities are almost constant and equal to 4.1 *cm/s*, though the aspect ratio of the channel is not constant. This is due to the fact that velocity profile in each experiment also changes due to the different sediments in suspension. The results therefore show that the bed shear velocity of an open channel is not only a function of the channel aspect ratio, but the location of the maximum velocity is also important. This fact is successfully captured by the present model. It can therefore be concluded that the present model can be extended to other flow conditions if the location of maximum velocity is known. The model can be effectively used for sidewall correction procedure in flume studies, because normally in these studies some experimental information on flow are available which can help estimating $\varepsilon$. In cases where the pattern of the velocity distribution is different, for example in trapezoidal or vegetated channels, the idea of the current research is still applicable with modifying the boundaries of $A_{wmin}$ and $A_{bmin}$. The equations for P1 and P2 might also be changed based on the velocity distribution.

Table 3. Comparison of present model for bed shear velocities (cm/s) with experimental data from Coleman (1986) for flow with suspended sediment

| Run | $M_s$ | $\varphi$ | h | b/h | h-$\varepsilon$ | $u_{*b,m}$ | $u_{*b,21}$ | $u_{*b,G\&J}$ |
|---|---|---|---|---|---|---|---|---|
|  | Kg | ×10$^{-4}$ | cm |  | cm |  | cm/s |  |
| 1 | 0 | 0 | 17.2 | 2.070 | 13.2 | 4.1 | 4.14 | 4.14 |
| 2 | 0.91 | 2.12 | 17.1 | 2.082 | 12.0 | 4.1 | 4.09 | 4.14 |
| 3 | 1.82 | 3.87 | 17.2 | 2.070 | 11.9 | 4.1 | 4.09 | 4.14 |
| 4 | 2.73 | 5.87 | 17.1 | 2.082 | 12.8 | 4.1 | 4.12 | 4.14 |
| 5 | 3.64 | 8.03 | 17.1 | 2.082 | 12.2 | 4.1 | 4.10 | 4.14 |
| 6 | 4.54 | 9.76 | 17.0 | 2.094 | 11.9 | 4.1 | 4.08 | 4.13 |
| 7 | 5.45 | 12.22 | 17.1 | 2.082 | 12.0 | 4.1 | 4.09 | 4.14 |
| 8 | 6.36 | 14.37 | 17.3 | 2.058 | 13.0 | 4.1 | 4.14 | 4.15 |
| 9 | 7.27 | 16.29 | 17.2 | 2.070 | 13.6 | 4.1 | 4.16 | 4.14 |
| 10 | 8.18 | 18.56 | 17.1 | 2.082 | 13.4 | 4.1 | 4.15 | 4.14 |
| 11 | 9.09 | 20.34 | 16.9 | 2.107 | 12.2 | 4.1 | 4.10 | 4.13 |
| 12 | 10 | 21.69 | 17.3 | 2.058 | 13.4 | 4.1 | 4.15 | 4.15 |
| 13 | 10.91 | 23.92 | 17.1 | 2.082 | 12.2 | 4.1 | 4.10 | 4.14 |
| 14 | 11.82 | 25.64 | 17.1 | 2.082 | 12.2 | 4.1 | 4.10 | 4.14 |
| 15 | 12.73 | 21.51 | 17.1 | 2.082 | 11.9 | 4.1 | 4.09 | 4.14 |
| 16 | 13.64 | 28.39 | 17.1 | 2.082 | 11.2 | 4.1 | 4.06 | 4.14 |
| 17 | 14.54 | 23.15 | 17.1 | 2.082 | 13.1 | 4.1 | 4.14 | 4.14 |

| 18 | 15.45 | 28.56 | 17.2 | 2.070 | 13.1 | 4.1 | 4.14 | 4.14 |
| 19 | 16.36 | 30.16 | 17.0 | 2.094 | 13.1 | 4.1 | 4.13 | 4.13 |
| 20 | 17.27 | 31.24 | 17.0 | 2.094 | 12.6 | 4.1 | 4.11 | 4.13 |
| 21 | 0 | 0 | 16.9 | 2.107 | 12.8 | 4.1 | 4.12 | 4.13 |
| 22 | 0.91 | 1.80 | 17.0 | 2.094 | 12.2 | 4.1 | 4.10 | 4.13 |
| 23 | 1.82 | 3.45 | 17.0 | 2.094 | 11.9 | 4.1 | 4.08 | 4.13 |
| 24 | 2.73 | 5.27 | 16.9 | 2.107 | 12.2 | 4.1 | 4.10 | 4.13 |
| 25 | 3.64 | 7.08 | 16.7 | 2.132 | 10.4 | 4.0 | 4.01 | 4.12 |
| 26 | 4.54 | 8.84 | 17.1 | 2.082 | 13.0 | 4.1 | 4.13 | 4.14 |
| 27 | 5.45 | 10.37 | 16.8 | 2.119 | 12.2 | 4.1 | 4.09 | 4.12 |
| 28 | 6.36 | 12.00 | 17.0 | 2.094 | 12.2 | 4.1 | 4.10 | 4.13 |
| 29 | 7.27 | 13.65 | 16.8 | 2.119 | 13.0 | 4.0 | 4.13 | 4.12 |
| 30 | 8.18 | 15.00 | 16.8 | 2.119 | 13.7 | 4.1 | 4.16 | 4.12 |
| 31 | 9.09 | 15.93 | 17.2 | 2.070 | 11.9 | 4.1 | 4.09 | 4.14 |
| 32 | 0 | 0 | 17.3 | 2.058 | 13.1 | 4.1 | 4.14 | 4.15 |
| 33 | 0.91 | 0.42 | 17.4 | 2.046 | 12.8 | 4.1 | 4.13 | 4.15 |
| 34 | 1.82 | 0.71 | 17.2 | 2.070 | 13.1 | 4.1 | 4.14 | 4.14 |
| 35 | 2.73 | 1.16 | 17.2 | 2.070 | 12.2 | 4.1 | 4.10 | 4.14 |
| 36 | 3.64 | 1.89 | 17.1 | 2.082 | 12.2 | 4.1 | 4.10 | 4.14 |
| 37 | 4.54 | 2.30 | 16.7 | 2.132 | 11.9 | 4.1 | 4.08 | 4.12 |

It should be noted that the present method can be used as far as the location of maximum velocity, $\varepsilon$, is known. $\varepsilon$ may be available from experiment or it can be calculated based on other hydraulic parameters of open channel flow using a graph presented by Chiu and Hsu (2006) [26].

While the equations proposed in the present research are derived for rectangular open channels, the proposed method of dividing the channel cross section into subsections can be used for channels with other cross-sectional shapes. Knight et al. (2007) [56] and Knight and Sterling (2000) [57] provided interesting information about the secondary current structures and velocity distribution in trapezoidal channels and circular partially full channels, respectively. Moreover Xang and Knight (2008) [58] and Yang et al. (2007) [59] studied the flow pattern in compound open channels with regular flow and vegetated flow, respectively. These works can be handy for researchers who might be interested in applying our method on other cross-sectional shapes or other flow regimes.

## 6. Conclusion

Determining the percentage of the shear force acting on the walls, $\%SF_w$ or bed, $\%SF_b$ in open channels has been the subject of research for several decades. The complex flow structure and formation of secondary flows make has been always a challenge in studying open channels. There is a series of methods for calculating *%SF$_w$* and *%SF$_b$* based on balancing the gravitational force in different regions of the cross section with the shear force of the channel bed and walls. In these methods by assuming zero-shear division lines channel cross section is divided into subsections. It is shown in this paper that a real zero-shear line does not exist in general.

In the present paper the channel cross section is divided into parts using channel bisectors, at which shear force due to secondary flows is zero, and isovel orthogonal trajectories, at which shear force due to velocity gradients is zero. The weight component of water in each of these subsections should be balanced either with bed or wall shear force. Based on these subsections the minimum and maximum possible shear force on walls and bed are calculated. To the best of our knowledge our method is the only one providing this range. The experimental data presented by various researchers lies well between the suggested maximum and minimum values. Considering the secondary flow structure, the share of the wall and bed from different subsections are calculated. An equation is finally derived for *%SF$_w$* and *%SF$_b$*. Calculated wall and bed shear forces are in good agreement with experimental data with an average relative error less than 5%.

The presented method can be used where the location of the maximum velocity, $\varepsilon$, is known. $\varepsilon$ could be available from experiment or in case of no velocity data it can be calculated directly from empirical dip equations or indirectly from velocity profile laws. The dependence of our equations on $\varepsilon$, makes it applicable to wide range of steady-uniform open channel flows. We also applied our equations to flows with suspended sediment, at which the location of the maximum velocity was known, with good agreement. While our equations work only for rectangular channels the general concept is applicable on channels with different cross-sections

## 7. Notations

*A*= area of the control surface

$A_{bmin}$, $A_{wmin}$, $A_1$, $A_2$= subsections on channel cross section

*Q*= discharge

$R^2$= determination coefficient

*RD*= relative difference

*S*= channel slope

$\%SF_b$= percentage of the total shear force acting on the bed

$\%SF_w$= percentage of the total shear force acting on the walls

**V**= velocity component of flow on y-z plane

*b*= channel width

*ε*= depth of the maximum velocity

*h*= water depth

*g*= gravitational acceleration

*l*= length along the boundary of surface

**n**= unit normal vector pointing outside of the control volume

$\eta_1$ and $\eta_2$= parameters related to $A_1$ and $A_2$, respectively

*ρ*= density

*φ*= volumetric suspended sediment concentration

$\overline{\tau}_b, \overline{\tau}_w$= mean bed and wall shear stress, respectively

*u, v, w*= components of mean velocity in x, y, and z directions, respectively

$u_{*b,G\&J}$ = shear velocity predicted by Gou and Julien (2005)

$u_{*b,m}$ = experimentally measured shear velocity

*v*= kinematic viscosity

$v_t$= turbulence viscosity

$u_{*b}$ = mean shear velocity on bed

$\tau_{nx}$ = shear stress in the flow direction x applying on the plane perpendicular to n

x = longitudinal coordinate

y = vertical coordinate

z = spanwise coordinate

## 9. Biographies

**Sasan Tavakkol** is a PhD candidate at University of Southern California (USC), Los Angeles, USA. His PhD research is on developing interactive real-time models for simulation and visualization of coastal waves and to employ such a model along with crowdsourced data in disaster situation to increase situational awareness. He received his second M.Sc. in computer science from USC where his research project was on prioritizing crowdsourced data. Sasan obtained his B.Sc. and first M.Sc. from Amirkabir University of Technology with the highest GPA. His research topics there, were on smoothed particle hydrodynamics (SPH) and open channel flows.

**Amir Reza Zarrati** is a professor of Hydraulic Engineering at the Department of Civil and Environmental Engineering, Amirkabir University of Technology Tehran, Iran. He has more than 20 years of teaching and research experience. His main area of research is numerical and physical modelling of flow behaviour, such as in hydraulic structures and rivers. He is also interested in scouring phenomenon and methods of its control. Dr. Zarrati has also a long collaboration with industry and has been senior consultant in a number of large dams in the country. He is a member of Iranian Hydraulic Association (IHA) and International Association of Hydraulic Research (IAHR).